\documentclass[12pt]{article}
\usepackage{natbib}
\usepackage{graphicx}
\usepackage{epsfig}
\usepackage{amssymb}
\begin{document}

\baselineskip24pt

\centerline{\bf Planetary Chaos and the (In)stability of Hungaria Asteroids} 

\bigskip
\centerline{Matija \' Cuk$^1$, David Nesvorn\' y$^2$}

\bigskip

\centerline{$^1$Carl Sagan Center, SETI Institute}
\centerline{189 North Bernardo Avenue, Mountain View, CA 94043}

\centerline{$^3$Southwest Research Institute}
\centerline{1050 Walnut St, Suite 400, Boulder, CO 80302}

\bigskip

\centerline{E-mail: cuk@seti.org}

\vspace{24pt}
\centerline{Re-submitted to Icarus}
\centerline{April 4$^{\rm th}$ 2017}

\vspace{24pt}

\centerline{Manuscript Pages: 27}

\centerline{Figures: 5}

\centerline{Tables: 1}
\newpage

Proposed Running Head: (In)stability of Hungaria Asteroids

\vspace{48pt}

Editorial Correspondence to:

Matija \' Cuk

Carl Sagan Center

SETI Institute

189 N Bernardo Ave

Mountain View, CA 94043

Phone: 650-810-0210

Fax: 650-961-7099

E-mail: mcuk@seti.org

\newpage

\noindent {ABSTRACT: The Hungaria asteroid group is located interior to the main asteroid belt, with semimajor axes between 1.8 and 2 AU, low eccentricities and inclinations of 16-35 degrees. Recently, it has been proposed that Hungaria asteroids are a secularly declining population that may be related to the Late Heavy Bombardment (LHB) impactors \citep{cuk12, bot12}. While \citet{cuk12} and \citet{bot12} have reproduced a Hungaria-like population that declined exponentially, the real Hungarias were never confirmed to be unstable to the same degree. Here we find that the stability of Hungarias is strongly dependent on the evolution of the eccentricity of Mars, which is chaotic and unpredictable on Gyr timescales. We find that the high Martian eccentricity chiefly affects Hungarias through close approaches with Mars, rather than planetary secular modes. However, current minimum perihelia of Hungarias (over Myr timescales) are not diagnostic of their long-term stability due to a number of secular and mean motion resonances affecting the Hungaria region \citep{mil10}. We conclude that planetary chaos makes it impossible to determine the effective lifetimes of observed Hungarias. Furthermore, long-term changes of Martian eccentricity could lead to variable Hungaria loss over time. We speculate that some of the most stable Hungarias may have been placed in their present orbit when the eccentricity of Mars was significantly higher than today.}


\newpage 

\section{Introduction}

Hungarias are a dynamical group of asteroids interior to the asteroid belt but exterior to the orbit of Mars (in the 1.8-2 AU range). Most stable Hungarias have high inclinations (16-35$^{\circ}$) and low eccentricities ($<0.1$). Hungarias are bracketed in inclination by multiple secular resonances, and are separated from the main asteroid belt by the $\nu_6$ secular resonance and the 4:1 mean-motion resonance with Jupiter. Hungarias have a less well-defined inner boundary that is being enforced by close encounters with Mars \citep{war09, mil10}. While they all inhabit the same island of relative dynamical stability, Hungarias are not compositionally uniform, with most common asteroid types being S and E \citep{war09, luc17}. A significant fraction of Hungarias belong to an E-type genetic family centered on 434 Hungaria \citep{war09}, which has been proposed as the main source of aubrite meteorites \citep{gaf92, cuk14}.

Unlike the asteroids in the main belt, Hungarias are thought not to be stable over the age of the Solar System, but are constantly escaping into the Mars-crossing region \citep{mil10, mce10}. The present-day reverse evolution of Mars-crossers into Hungarias has been found insufficient for replenishing the Hungaria population \citep{mce10}. The implication is that the proto-Hungarias were orders of magnitude more numerous in the early Solar System, and may have been the main source of the Late Heavy Bombardment \citep{cuk12, bot12}. Hungarias are proposed to be depleted remnants of primordial Mars-crossers \citep{cuk12}, or survivors from the extinct innermost part of main asteroid belt perturbed by late planetary migration \citep{bot12}. \citet{cuk12} showed that a primordial population consisting solely from Mars-crossers can give rise to a quasi-stable Hungaria-like population, which eventually has a half-life of 600 Myr. \citet{bot12} started with an initially dynamically cold ``E-belt'' that was perturbed by late giant planet migration, but the end results was similar in terms of creating quasi-stable Hungarias that slowly declined over next few Gyr \citep{nes16}. 

Given that the idea of a exponentially declining Hungaria population is established in the literature based on modeling of synthetic bodies, we wanted to test this concept on real asteroids. Previously, \citet{mig98} have reported a 960-Myr half-life for largest Hungarias, which may be significantly different from \citet{cuk12} prediction of 600 Myr, while \citet{gal14} found 5\% loss over 100 Myr, implying half-life beyond 1 Gyr. Additionally, it would be very interesting to determine if there is a difference in dynamical stability between different spectroscopic types of Hungarias. Currently, 80\% of background Hungarias (i.e. excluding the E-type family centered on 434 Hungaria) are S-type asteroids \citep{luc17}. While the E-type Hungarias are very likely enstatite achondrites \citep[i.e., aubrites ][] {gaf92, cuk14}, S-type Hungarias have been suggested to contain more primitive achondrites than ordinary chondrites than the main-belt S-type asteroids \citep{luc17}. Additionally, some of the few rare A-type (olivine-rich) asteroids are found among Hungarias. Different dynamical half-lives may indicate that the some of the spectral types may have been more of less common in the past, with implications for the primordial distribution of small bodies in the Solar System.


\bigskip

\section{Billion-Year Numerical Experiments}

\bigskip

Before moving on to purely gravitational dynamics of Hungarias, we must address the radiative Yarkovsky and YORP effects \citep[][ and references therein]{rub95, rub00, bot06}. Because none of the Hungarias are very large, almost all of them should experience some semimajor axis mobility due to the Yarkovsky effect, which is dependent on the objects' rotations, which are themselves affected by the YORP effect. An ideal simulation of Hungarias' stability would (in addition to perturbations from all eight planets) include Yarkovsky and YORP effects and close approaches with Mars (when needed). Another requirement is more subtle: we need all of the particles to interact with planets on the same orbits. Integrators such as {\sc swift}-rmvs3 \citep[and its Yarkovsky variant, {\sc swift}-rmvsy; ][]{lev94, bro06} which are both efficient and handle close encounters between planets and test particles well, have a peculiarity that the evolution of planetary orbits depends on the motion of otherwise massless particles. This is because the special handling of close approaches introduces time-step changes, the timing of which depends on what the test particles are doing. A large set of Hungarias (and their clones) would need to be integrated on many CPUs, and given the serial nature of {\sc swift} (constrained by the nature of a n-body problem) every node would in practice integrate a different version of the planetary system. Our integrations are not optimized to follow planetary orbits as precisely as possible, but even state-of-the-art numerical experiments cannot predict planetary orbits more than 50-60 Myr in the future \citep{las11a, las11b}. On the other hand, {\sc swift}-rmvs4, while not including Yarkovsky effect, is designed to integrate close approaches without having the test particles affect planetary orbits in any way, so that all nodes will experience the same planetary system. A more primitive symplectic integrator {\sc simpl} that one of the authors (M.\' C.) used in \citet{cuk15} can include arbitrary forces like the Yarkovsky effect, but does not include close encounters, which makes it unsuited to this study but also happens to eliminate the problem of planetary divergence.

In order to be able control for the evolution of planetary orbits, we have decided to primarily use {\sc swift}-rmvs4 for our numerical experiments. We wanted to retain the ability to accurately integrate close approaches with Mars, and decided that the accurate treatment of the Yarkovsky effect is not important to the same degree. To partially account for the Yarkovsky-related semimajor axis drift, we cloned each body 21 times (including the nominal orbit) and spread the clones within the range of semimajor axis expected from Yarkovsky drift in 100 Myr, using estimates based on \citet{bot06} (Table \ref{table}). We generally find little correlation between fates of neighboring particles so we do not think that Yarkovsky-driven semimajor axis spreading is a major effect on the stability of largest Hungarias on the sub-Gyr timescale.

\begin{figure*}[h]
\includegraphics[]{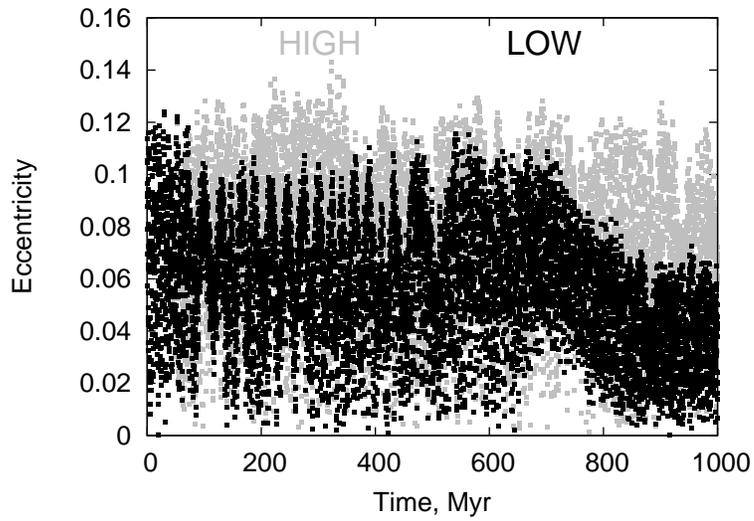}  
\caption{The evolution of Martian eccentricity over next 1 Gyr in our two cases of the dynamical evolution of the Solar System, {\sc high} and {\sc low}. The two simulations differ only by a small shift in the initial position of the planet Mercury.} 
\label{mars}
\end{figure*}

The importance of variations in Martian eccentricity for the stability of Hungarias was first indicated by inconsistent results we were getting for their stability when using slightly different initial conditions or different integrators. Ever since \citet{las94} it was known that Martian eccentricity can experience large variation over the age of the Solar System, and \citet{cuk15} recently found a major effect of the variation in Martian eccentricity on the dynamics of Mars Trojans. This suggested that we may need to separately explore Hungarias' stability in different ``parallel Solar systems'' with different histories of Martian eccentricity. We first set up ten different simulations of the planetary system alone using {\sc swift}-rmvs4, with identical initial conditions\footnote{Just like the nominal orbits of the Hungarias, planetary initial conditions were based on vectors obtained from JPL's HORIZONS system.}, except that in each next integration (starting with the second) Mercury was $10^{-7}$~AU further from the Sun in the x-direction. After 1 Gyr, we compared the behavior of Martian eccentricity in all ten simulations, and chose two of them as initial conditions for our two cases of Hungaria stability, which we will term {\sc high} and {\sc low} (Fig. \ref{mars}).

\begin{figure*}[h]
\includegraphics[]{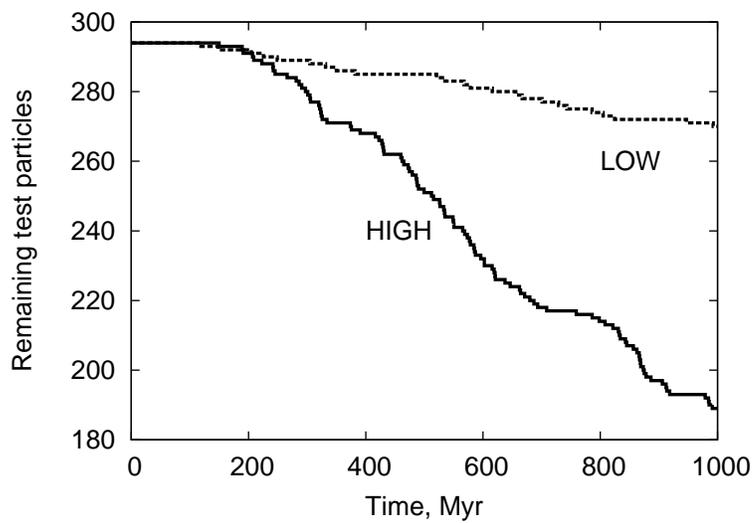}  
\caption{Number of surviving Hungaria clones in our two cases, {\sc high} and {\sc low}, out of original 294 particles. The initial conditions for the nominal orbit of each of the fourteen Hungarias, plus additional twenty clones per asteroid (see Table \ref{table}), were the same in the {\sc high} and {\sc low} simulations, and the two cases differed only by planetary initial conditions.} 
\label{histo}
\end{figure*}

We then ran 1 Gyr simulations of the clones of 14 different, mostly brightest, Hungarias (Table \ref{table}; we also included 5639 Cuk as a less stable case). We ran the same 21 initial conditions in both the {\sc high} and {\sc low} Martian eccentricity histories. As Table \ref{table} and Fig. \ref{histo}, there is a substantial difference between the number of surviving clones between {\sc high} and {\sc low} histories (which differ only in Mercury's initial position but have the same starting vectors for Hungaria clones). The difference is largest for 1509 Esclangona and 5639 Cuk, but it is also pronounced for six other Hungarias (Table \ref{table}). The remaining cases appear more stable, but that does not rule out instability for even higher values of Martian eccentricity. 

A clear implication of this numerical experiment is that we cannot assign a stability timescale to Hungarias as a group without specifying the future behavior of the Martian eccentricity on Gyr timescales. Since we cannot know the future evolution of inner planets' orbits on Gyr timescale with any confidence \citep{las11a, las11b}, there cannot be a single number quantifying dynamical lifetime for most Hungarias. We will address some questions of relevant dynamical mechanisms in the next Section, and discuss implications for the origin of Hungarias in the final Section.

\bigskip

\section{The Dynamical Mechanism of Mars-Hungaria Connection}

\bigskip

Our finding that Martian eccentricity affects Hungarias leads to an obvious question: are Hungarias being excited {\it secularly} by a more eccentric Mars, or do the larger aphelion distances of an eccentric Mars allow for close encounters with asteroids which were out of Mars's reach when the planet had lower eccentricity? Long-term increase in Martian eccentricity happens due to the slow increase of power in the $g_4$ eigenmode of the planetary system \citep{md99}. Is it possible that this strengthening of $g_4$ also affects Hungarias, making them also more eccentric and less stable?

\begin{figure*}[h]
\includegraphics[]{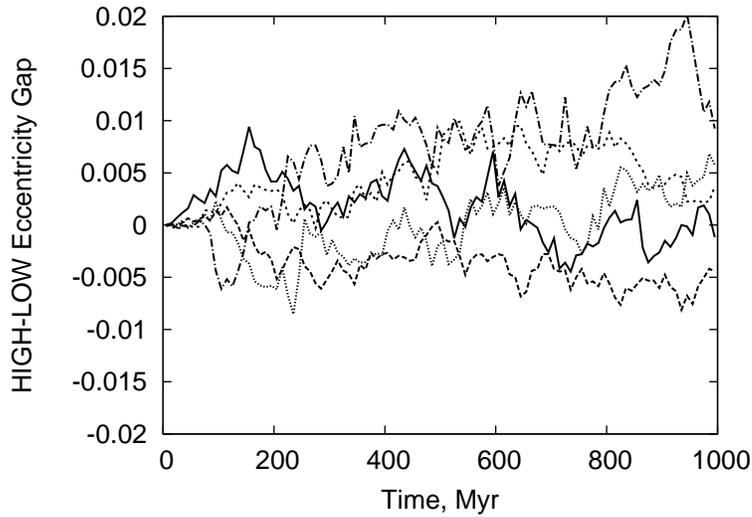}  
\caption{Difference between the average eccentricity of the clones in {\sc high} and {\sc low} simulations for four relatively stable Hungarias: 424 Hungaria (solid line), 1019 Strackea (long dashes), 1453 Fennia (short dashes), 3266 Bernardus (dotted line) and 5806 Archieroy (dash-dot line). Only clones that did not experience any close encounters in either {\sc high} or {\sc low} simulation were included into the average, and the above-listed asteroids had 17, 20, 19, 15 and 21 such zero-encounter clones, respectively. These results suggest that there is no large systematic eccentricity difference between clones in {\sc high} and {\sc low} simulation for different Hungarias, at least in the cases when the clones are only affected by secular planetary perturbations.} 
\label{ecc}
\end{figure*}

We attempted to address this issue without having to construct a full secular model of Hungarias' eigenmodes, by empirically comparing the behavior of clones in the {\sc high} and {\sc low} simulations. However, to be sure we are measuring only secular effects, we had to exclude clones that had close encounters with Mars (which are logged by {\sc swift}). We therefore focused on objects that had most clones without any close encounters in either {\sc high} or {\sc low} simulations: 434 Hungaria, 1019 Strackea, 1453 Fennia, 3266 Bernardus, and 5806 Archieroy with 17, 20, 19 and 15 zero-encounter clones each, respectively. We then averaged the eccentricity for all the suitable clones of each Hungaria over every 10 Myr in both {\sc high} and {\sc low} simulations. Figure \ref{ecc} plots the difference between the averaged values over all such clones in {\sc high} and {\sc low} simulations for each Hungaria. We cannot see any trend common to all four sets of clones, with the results shown in Fig. \ref{ecc} being consistent with a random drift between {\sc high} and {\sc low} simulations, with no preferentially higher eccentricity for Hungarias' (non-encounter) clones in the {\sc high} simulation. Large variation seen for Archieroy is the results of almost all of its clones experiencing substantial eccentricity growth over 1 Gyr, as the current low eccentricity ($e=0.03$) appears to be ephemeral. Large changes in eccentricity also lead to larger absolute variations between {\sc low} and {\sc high} simulations, without a necessarily meaningful trend. While we may be missing some of the secular effects by excluding the clones that become eccentric enough to experience close encounters, we can say that this is not a large systematic effect common to all Hungarias. This is in contrast with Mars Trojans, which seem to have significant components of their eccentricity forced by Mars \citep{cuk15}. We therefore conclude that the likely mechanism by which Mars interacts with the Hungarias is primarily through close encounters, which is consistent with our qualitative inspection of the orbital histories of unstable clones.

\begin{figure*}[h]
\includegraphics[]{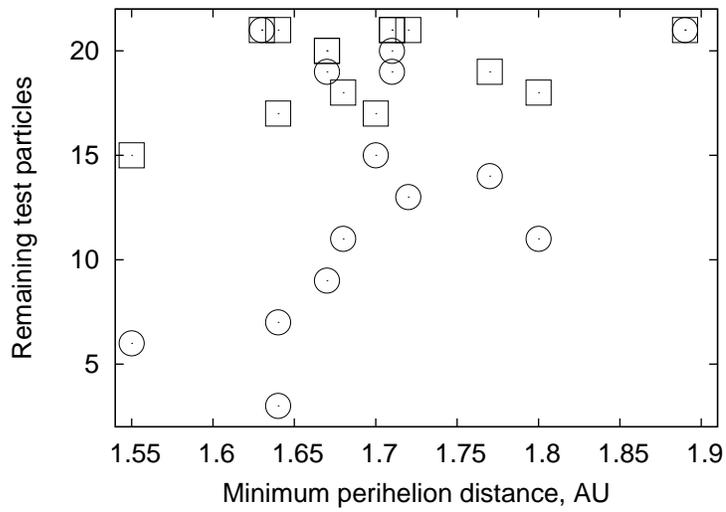}  
\caption{Number of surving clones (out of total of 21) in {\sc high} (circles) and {\sc low} (squares) cases for each Hungaria, ploted againts its aproximate lowest perihelion distance over a 10 Myr simulation. It is evident that the stability of Hungarias is not determined by their short-term perihelion distance alone.} 
\label{peri}
\end{figure*}

As we think that close encounters are a primary mechanism by which a more eccentric Mars destabilizes Hungarias, one possible implication would be that a Hungaria's stability against Martian encounters is determined only by the asteroid's perihelion distance. It that were true, we would only need to determine a Hungaria's minimum perihelion distance in a (relatively) short-term numerical simulation, and that would give us a direct measure of its susceptibility to destabilization by Mars. In other words, as Mars becomes more eccentric in our {\sc high} simulation, this simple picture would require Hungarias to become destabilized in ascending order of their perihelion distances. Fig. \ref{peri} tests this assumption by plotting number of surviving clones in {\sc high} and {\sc low} cases against each Hungaria's minimum perihelion over next 10 Myr. There is no clear correlation between the current minimum perihelion distance and stability, indicating that the long-term dynamics of individual Hungarias is determined by other factors.  

\begin{figure*}[h]
\includegraphics[]{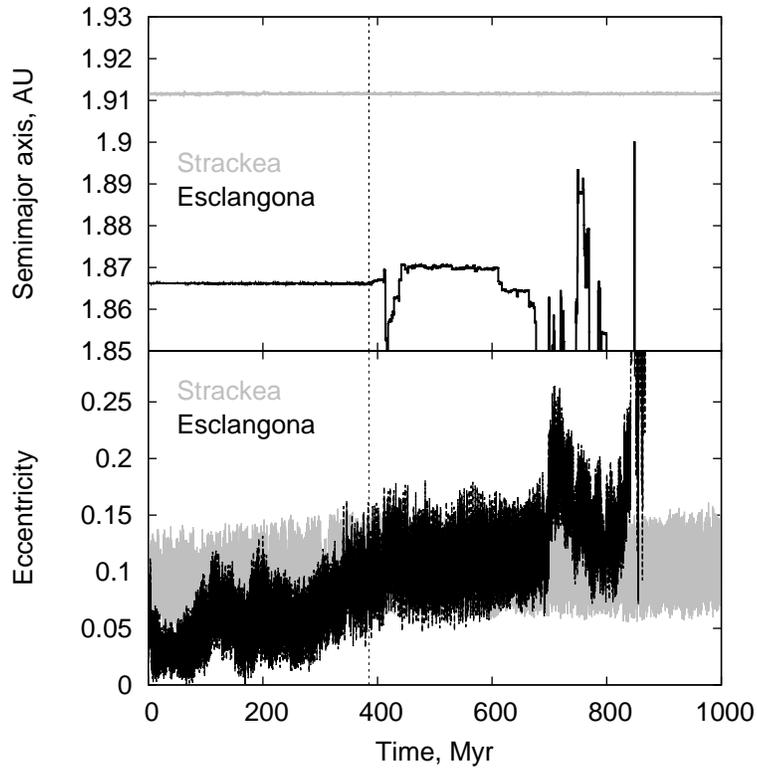}  
\caption{Semimajor axis (top panel) and eccentricity (lower panel) evolution of ``nominal" (i.e. unshifted in $a$) 1019 Strackea (grey lines) and 1509 Esclangona (black lines) in the course of the {\sc high} simulation. While the two objects have similar minimum perihelion distances at the present epoch, their long-term dynamics are very different, even before Esclangona starts having close encounters with Mars at 385 Myr (vertical dashed line).} 
\label{a_and_e}
\end{figure*}

A good example of how seemingly unpredictable the behavior of individual Hungarias can be is the contrast between 1019 Strackea and 1509 Esclangona. While they have similar minimum perihelia over 10 Myr (1.63~AU vs 1.64~AU, respectively) all of Strackea's clones are stable even in the {\sc high} simulation, while more than half of Esclangona's clones are destabilized. Fig. \ref{a_and_e} plots the semimajor axis and eccentricity of the nominal solution for Strackea and Esclangona in the {\sc high} simulation. While the eccentricity of Strackea remains stable and low, eccentricity of Esclangona is significantly variable on 100~Myr timescales, and eventually becomes high enough that it drives Esclangona into close encounters with Mars at 385~Myr (vertical line). Note that the semimajor axis of Esclangona is completely stable up to that event, indicating that the perturbations that Esclangona is experiencing are secular \citep[Esclangona is not also near any mean-motion resonances; ][]{mce10}. We note that Esclangona and its clones also exhibit variable eccentricity in the {\sc low} simulation, with the difference that eccentricities of Esclangona up to 0.2 do not lead to close encounters with Mars when the Martian eccentricity is low\footnote{Hungarias do not start encountering Mars immediately when there is a formal overlap in heliocentric distance; this is because inclined Hungarias tend to experience maximum eccentricity when the longitude of perihelion is perpendicular the line of nodes \citep{koz62}.}. 

Results of \citet{mil10} plotted in their Fig. 11 indicate that Esclangona may be affected by $g-s-g_5+s_6$ secular resonance; the anti-correlated behavior of its eccentricity and inclination is consistent with that hypothesis. Large number of different (relatively weak) secular and mean-motion resonances in the Hungaria region \citep{war09, mil10, mce10} make it hard to estimate a Hungaria's stability without a detailed examination. Once we introduce the Yarkovsky effect, it becomes even harder to predict the future behavior of any Hungaria without doing a long term integration involving multiple clones, and even then we can only expect probabilistic results. However, we can still make some conclusions about the dynamics of Hungarias as a group, which is the topic of the final section. 

To summarize the results of this section, we found that Martian eccentricity does not directly affect the secular behavior of the Hungarias, leaving close encounters as the main mechanism of Hungaria-Mars interaction. Therefore, the larger aphelion distance of Mars in the {\sc high} simulation is the main driver of addition destabilization of Hungarias. We also find that some Hungarias are apparently affected by secular resonances and near-resonances \citep[in agreement with][] {mil10}, which make it hard to predict those Hungarias' Gyr-scale stablity from their orbits integrated over Myr scales. There is no contradiction between these findings, as relevant secular resonances in the Hungaria region do not involve Mars's main eccentricity eigenmode $g4$ \citep[Fig. 11 in ][]{mil10} and are therefore expected to have the same effect in {\sc high} and {\sc low} simulations.   

\bigskip

\section{Discussion and Conclusions}

\bigskip

The results of our 1 Gyr simulations show that the Hungaria stability is sensitively dependent on Martian eccentricity, which is itself strongly affected by the long-term planetary chaos. We find that Hungarias do not have a dynamical half-life that can be separated from the behavior of Martian eccentricity. This is different from the view proposed by \citet{cuk12} and \citet{bot12} that Hungarias can be considered an exponentially declining population, with a half-life of about 600 Myr. It appears that the conclusions of \citet{cuk12} are specific to their eccentricity history of Mars, while \citet{bot12} used {\sc swift}-rmvs3 which means that their test particles inhabited more than one ``parallel Solar System'' with different histories of Martian eccentricity. However, since \citet{bot12} \citep[as well as ][]{nes16} did not include Mercury, the orbit in Mars is their simulations is likely to be much less chaotic than in the real Solar System. In any case, it appears that the dynamics of Hungarias is more complex than was thought before, which complicates the estimates of proto-Hungaria population at the time of LHB that is central to the models of \citet{cuk12} and \citet{bot12}.

Beyond the dependence on Martian eccentricity, our sample of Hungarias is more stable than those produces by \citet{cuk12} and \citet{bot12}, as the overall half-life is significantly longer than 1 Gyr. To assess how ``typical'' our two histories are, we need to consult results of \citet{las08}, who calculated probabilities of planets reaching different eccentricities and inclinations over different periods of time. \citet{las08} find that Mars has the probability of about 30\% to reach the eccentricity of $0.15$ in 1 Gyr. Since the highest eccentricity Mars reached in any of our 10 planetary simulations is 0.142 (in the simulation that was chosen as the {\sc high} case), it appears that our simulations somewhat underestimated the high end of Martian eccentricity variation. One possible contributing factor may be that our integrations do not include the effects of General Relativity on the orbit of Mercury, which are known to be important for the long-term chaos among the inner planets \citep{las08}. Inclusion of General Relativity lowers the long-term eccentricity of Mercury, and the eccentricities of Mercury of Mars are known to be anti-correlated on long timescales \citep{las94}. Therefore we cannot claim that relatively high stability of some Hungarias in our {\sc high} simulation indicates that they would be as stable in the real Solar System over 4.5 Gyr.

Another possibility for explaining the relatively long stability timescales of our Hungaria clones relative to test particles used by \citet{cuk12} and \citet{bot12} is that the destabilization of Hungarias is not a simple exponential decay. \citet{min10} have found that real Solar System populations have long-lived ``tails'', as remaining bodies tend to be in very stable dynamical islands. This is an interesting possibility that needs to be investigated, but it does not contradict our conclusion that the observed population of Hungarias is not a direct and unequivocal indicator of a large primordial population envisaged by \citet{cuk12} and \citet{bot12}. 

Even within our small sample, it appears that both S and E/X asteroid types contain both very stable and relatively unstable representatives among Hungarias. While a larger sample may be able to reveal more subtle trends in the stability of different spectral classes, we find no evidence so far of connection between dynamics and physical properties among largest Hungarias.

Taking a longer-term view, according to \citet{las08} Mars has 50\% chance of reaching eccentricity of 0.16 over 5~Gyr. So the fact that some Hungarias are currently safely outside of Mars's reach does not imply that that has always been the case. Therefore, it is still possible that Hungarias have been planted into their present orbits by past close encounters with Mars, as suggested by \citet{cuk12} and \citet{bot12}. In that case, it appears likely that the eccentricity of Mars may have been higher in the early Solar System than it is now. This may have led to at least somewhat non-uniform decrease of impact on the Moon and terrestrial planets during the so-called LHB, assuming that the proto-Hungarias were a major source of the impactor flux. However, the fact that it is impossible to uniquely reconstruct the past population of Hungarias (as it is a stochastic problem) opens the possibility for the rival hypothesis that Hungarias were planted in their present orbits by precesses other than Martian encounters. For example, processes thought to have placed main belt asteroids onto their present orbits include now-extinct planetary embryos and/or dramatic planetary migration \citep{mor15}. If the Hungarias were placed onto their orbits the same way as the main-belt asteroids, the Hungarias did not descend from a massive early population of Mars-crossers. In that case, past population of Hungarias was likely less numerous and more stable than \citet{cuk12} and \citet{bot12} have proposed, and proto-Hungarias would not have been a major source of the LHB. More work on the stability of Hungarias is clearly needed, including in cases when the eccentricity of Mars was close to its maximum expected value over 5 Gyr. If almost all Hungarias become destabilized at this point, it is plausible that they may have also been planted by Mars and that the scenarios of \citet{cuk12} and \citet{bot12} may be broadly correct. Otherwise, if many Hungarias are impervious to Martian perturbation even at Martian eccentricities of 0.16 or so, it would appear that Hungarias are similar to main-belt asteroids in being broadly stable, rather than being a true metastable remnant of terrestrial planet formation. 

The above discussion ignores the non-gravitational Yarkovsky effect, which is relevant when discussing the present population of Hungarias, all of which are definitely below 20~km in diameter. It is possible that Yarkovsky can ``scramble" the orbital distribution of Hungarias so that some less stable ones become more stable and vice versa. However, since the Yarkovsky effect is essentially random (and should vary over time due to the YORP effect) there is no reason to think that the dynamical characteristics of the whole Hungaria population will be irreversibly altered only through the Yarkovsky effect. Only extensive future numerical simulations that fully include both planetary chaos and the Yarkovsky effect will be able to resolve these issues. 

\vspace{48pt}

{\centerline{\bf ACKNOWLEDGMENTS}}

\bigskip

M\' C is supported by NASA's Planetary Geology and Geophysics (PGG) program, award number NNX12AO41G. DN's work is supported by NASA'a Solar System Exploration Research Virtual Institute (SSERVI). M \'C thanks Lucy Lim and Michael Lucas for their help in choosing Hungaria asteroids for this study. We thank Hal Levison for making the SWIFT-rmvs4 integrator available to the community. We also thank David Minton and an anonymous reviewer for their comments which have improved the paper.

\bibliographystyle{}

\begin{thebibliography}{1}





\bibitem[Bottke et al.(2006)]{bot06} Bottke, W.~F., Jr., Vokrouhlick{\'y}, D., Rubincam, D.~P., Nesvorn{\'y}, D.\ 2006.\ The Yarkovsky and Yorp effects: Implications for asteroid dynamics.\ Annual Review of Earth and Planetary Sciences 34, 157-191.



\bibitem[Bottke et al.(2012)]{bot12} Bottke, W.~F., Vokrouhlick{\'y}, D., Minton, D., Nesvorn{\'y}, D., Morbidelli, A., Brasser, R., Simonson, B., Levison, H.~F.\ 2012.\ An Archaean heavy bombardment from a destabilized extension of the asteroid belt.\ Nature 485, 78-81. 

\bibitem[Bro{\v z}(2006)]{bro06} Bro{\v z}, M.\ 2006.\ Yarkovsky effect and the dynamics of the Solar System.\ Ph.D.~Thesis, Charles University. Prague, Czech Republic.



\bibitem[\' Cuk(2012)]{cuk12} \' Cuk, M. 2012. Chronology and Sources of Lunar Impact Bombardment.\ Icarus 218, 69-79.

\bibitem[{\'C}uk et al.(2014)]{cuk14} {\'C}uk, M., Gladman, B.~J., Nesvorn{\'y}, D.\ 2014.\ Hungaria asteroid family as the source of aubrite meteorites.\ Icarus 239, 154-159. 

\bibitem[{\'C}uk et al.(2015)]{cuk15} {\'C}uk, M., Christou, A.~A., Hamilton, D.~P.\ 2015.\ Yarkovsky-driven spreading of the Eureka family of Mars Trojans.\ Icarus 252, 339-346. 



Solar System II 829-851.




\bibitem[Gaffey et al.(1992)]{gaf92} Gaffey, M.~J., Reed, K.~L., Kelley, M.~S.\ 1992.\ Relationship of E-type Apollo asteroid 3103 (1982 BB) to the enstatite achondrite meteorites and the Hungaria asteroids.\ Icarus 100, 95-109.

\bibitem[Galiazzo and Schwarz(2014)]{gal14} Galiazzo, M.~A., Schwarz, R.\ 2014.\ The Hungaria region as a possible source of Trojans and satellites in the inner Solar system.\ Monthly Notices of the Royal Astronomical Society 445, 3999-4007. 






\bibitem[Kozai(1962)]{koz62} Kozai, Y.\ 1962.\ Secular perturbations of asteroids with high inclination and eccentricity.\ The Astronomical Journal 67, 591. 

\bibitem[Laskar(1994)]{las94} Laskar, J.\ 1994.\ Large-scale chaos in the solar system.\ Astronomy and Astrophysics 287, L9-L12. 

\bibitem[Laskar(2008)]{las08} Laskar, J.\ 2008.\ Chaotic diffusion in the Solar System.\ Icarus 196, 1-15. 

\bibitem[Laskar et al.(2011a)]{las11a} Laskar, J., Fienga, A., Gastineau, M., Manche, H.\ 2011a.\ La2010: a new orbital solution for the long-term motion of the Earth.\ Astronomy and Astrophysics 532, A89. 

\bibitem[Laskar et al.(2011b)]{las11b} Laskar, J., Gastineau, M., Delisle, J.-B., Farr{\'e}s, A., Fienga, A.\ 2011b.\ Strong chaos induced by close encounters with Ceres and Vesta.\ Astronomy and Astrophysics 532, L4. 


\bibitem[Levison and Duncan(1994)]{lev94} Levison, H.~F., Duncan, M.~J.\ 1994.\ The long-term dynamical behavior of short-period comets.\ Icarus 108, 18-36. 

\bibitem[Lucas et al.(2017)]{luc17} Lucas, M.~P., Emery, J.~P., Pinilla-Alonso, N., Lindsay, S.~S., Lorenzi, V.\ 2017.\ Hungaria asteroid region telescopic spectral survey (HARTSS) I: Stony asteroids abundant in the Hungaria background population.\ Icarus, in press.

\bibitem[McEachern et al.(2010)]{mce10} McEachern, F.~M., {\'C}uk, M., Stewart, S.~T.\ 2010.\ Dynamical evolution of the Hungaria asteroids.\ Icarus 210, 644-654. 

\bibitem[Migliorini et al.(1998)]{mig98} Migliorini, F., Michel, P., Morbidelli, A., Nesvorny, D., Zappala, V.\ 1998.\ Origin of Multikilometer Earth- and Mars-Crossing Asteroids: A Quantitative Simulation.\ Science 281, 2022. 

\bibitem[Milani et al.(2010)]{mil10} Milani, A., Kne{\v z}evi{\'c}, Z., Novakovi{\'c}, B., Cellino, A.\ 2010.\ Dynamics of the Hungaria asteroids.\ Icarus 207, 769-794. 

\bibitem[Minton and Malhotra(2010)]{min10} Minton, D.~A., Malhotra, R.\ 2010.\ Dynamical erosion of the asteroid belt and implications for large impacts in the inner Solar System.\ Icarus 207, 744-757.


\bibitem[Morbidelli et al.(2015)]{mor15} Morbidelli, A., Walsh, K.~J., O'Brien, D.~P., Minton, D.~A., Bottke, W.~F.\ 2015.\ The Dynamical Evolution of the Asteroid Belt.\ Asteroids IV 493-507. 

\bibitem[Murray and Dermott(1999)]{md99} Murray, C. D., Dermott, S. F., 1999. Solar System Dynamics. Cambridge Univ. Press.

\bibitem[Nesvorn{\'y} et al.(2017)]{nes16} Nesvorn{\'y}, D., Roig, F., Bottke, W.~F.\ 2017.\ Modeling the Historical Flux of Planetary Impactors.\ The Astronomical Journal 153, 103. 

\bibitem[Rubincam(1995)]{rub95} Rubincam, D.~P.\ 1995.\ Asteroid orbit evolution due to thermal drag.\ Journal of Geophysical Research 100, 1585-1594. 

\bibitem[Rubincam(2000)]{rub00} Rubincam, D.~P.\ 2000.\ Radiative Spin-up and Spin-down of Small Asteroids.\ Icarus 148, 2-11. 



\bibitem[Warner et al.(2009)]{war09} Warner, B.~D., Harris, A.~W., Vokrouhlick{\'y}, D., Nesvorn{\'y}, D., Bottke, W.~F.\ 2009. \ Analysis of the Hungaria asteroid population.\ Icarus 204, 172-182. 




 

















\end{thebibliography}

\newpage

\begin{table}
\begin{center}
\caption{Results of our 1 Gyr simulations for 14 Hungarias. The albedos and magnitudes shown here were used for determining $\Delta a$. These vales were obtained from Jet Propulsion Laboratory's HORIZONS system in November 2014 and may be outdated (we used albedo of 0.3 when neither spectral class nor albedo were known). The spread in semimajor axes $\Delta a$ is determined after the fact from the simulation output (as the spread was introduced very roughly by changing the Hungaria's heliocentric ditance), and represents range from smallest to largest $a$ (with the nominal orbit in the center and other clones distributed uniformly troughout the range). N$_{\rm HIGH}$ and N$_{\rm LOW}$ give the number of clones still in the simulation (i.e. that have not been lost to collisions or ejection) at 1 Gyr.\label{table}}
\bigskip
\begin{tabular}{|l|c|c|c|c|c|c|}
\hline\hline
Asteroid & H mag. & Albedo & $\Delta a$ [AU] & N$_{\rm HIGH}$ & N$_{\rm LOW}$ & Class \\
\hline\hline
(434) Hungaria & 11.21 & 0.428 & 1.4$\times 10^{-3}$ & 19 & 21 & E\\
\hline
(3447) Burckhalter & 12.2 & (0.3) & 3.0$\times 10^{-3}$ & 11 & 18 & (u)\\
\hline
(1103) Sequoia & 12.25 & 0.4 & 2.1$\times 10^{-3}$ & 11 & 18 & E\\
\hline
(1025) Riema & 12.4 & 0.4 & 3.9$\times 10^{-3}$ & 15 & 17 & E\\
\hline
(1600) Vyssotsky & 12.5 & 0.25 & 2.8$\times 10^{-3}$ & 13 & 21 & A\\
\hline
(1453) Fennia & 12.5 & 0.2495 & 2.8$\times 10^{-3}$ & 20 & 21 & S\\
\hline
(1019) Strackea & 12.63 & 0.2265 & 2.3$\times 10^{-3}$ & 21 & 21 & S\\
\hline
(1509) Esclangona & 12.64 & 0.2327 & 3.0$\times 10^{-3}$ & 7 & 21 & S\\
\hline
(3940) Larion & 12.7 & (0.3) & 2.7$\times 10^{-3}$ & 14 & 19 & (u)\\
\hline
(3169) Ostro & 12.73 & 0.4 & 3.1$\times 10^{-3}$ & 9 & 20 & Xe\\
\hline
(3266) Bernardus & 12.8 & (0.3) & 7.2$\times 10^{-3}$ & 19 & 20 & (u)\\
\hline
(2001) Einstein & 12.85 & 0.4 & 5.7$\times 10^{-3}$ & 6 & 15 & E\\
\hline
(5806) Archieroy & 12.9 & (0.3) & 3.5$\times 10^{-3}$ & 21 & 21 & (u)\\
\hline
(5639) Cuk & 14.7 & (0.3) & 9.5$\times 10^{-3}$ & 3 & 17 & (u)\\
\hline

\end{tabular}
\end{center}
\end{table}

\end{document}